# Preface

Apparently, the detailed knowledge of the nature of resistive state of current-carrying superconducting thin films in magnetic field, external or self-current-induced, is critically important for designing superconducting devices like optical single-photon detectors and understanding some physical phenomena like ultra-fast magnetic vortices. Therefore, by initiative and kind permission of our colleague, Dr. A. G. Sivakov (ORCID [0009-0005-0349-0433](https://), [sivakov@ilt.kharkov.ua](mailto:sivakov@ilt.kharkov.ua)), who is the co-author of the paper below, we decided to remind the superconducting community about these rather simple but very informative and convincing experiments that reveal the magnetic vortex behavior within a non-equilibrium region nearby a phase slip line or an *S-N* boundary. The conclusion is that the vortices are always driven by solely superconducting component of the transport current.

The original version in Russian [*Fiz. Nizk. Temp.* 11(5), 547-551 (1985)] can be accessed through the journal webpage of Fizika Nizkikh Temperatur [https://fnt.ilt.kharkiv.ua/](https://fnt.ilt.kharkiv.ua/) at the website of B. Verkin Institute for Low Temperature Physics and Engineering (ILTPE) of the National Academy of Sciences of Ukraine (Ukrainian and English interface).

The corresponding English translation issued in Soviet Journal of Low Temperature Physics [*Sov. J. Low Temp. Phys.* 11(5), 300-302 (1985)] can be found at the website of AIP Low Temperature Physics via DOI: [https://doi.org/10.1063/10.0031298](https://doi.org/10.1063/10.0031298).

This postprint is a corrected version of the LTP paper. Having no access to the PDF, we restore it from the printed journal. The latter was scanned, OCRed and pre-formatted (A. V. Krevsun, [krevsun@ilt.kharkov](mailto:krevsun@ilt.kharkov)), then thoroughly revised to improve the translation (based on the Russian source), eliminate some misprints, add the lost footnote and finally format the layout (O. G. Turutanov, ORCID [0000-0002-8673-136X](https://), [turutanov@ilt.kharkov.ua](mailto:turutanov@ilt.kharkov.ua)).

We hope this paper is still of interest for those who study resistive states in superconductors and apply this knowledge for building thin-film devices for superconducting electronics.

# Vortex dynamics in the region of charge imbalance

A. G. Sivakov and V. G. Volotskaya

*Physicotechnical Institute of Low Temperatures, Academy of Sciences of the Ukrainian SSR, Kharkov*

The nature of vortex flows in the nonequilibrium region arising in the vicinity of phase-slip lines at the *S–N* boundary are investigated experimentally. It is shown that vortices continue to move when charge imbalance appears in a film that is in a dynamic mixed state. The vortex dynamics in this case is determined by the interaction of the vortices with only the superconducting component of the current. This statement is supported by a study of the dynamic mixed state in the nonequilibrium region near the *S–N* boundary.

Experimental investigations of the resistive state of wide films led to the discovery of features in the current-voltage (*I–V*) characteristics, which were not associated with the vortex flow. Phase-slip lines (PSLs) (Ref. 1), analogous in structure to the phase-slip centers (PSCs) in narrow superconducting channels,[2] were shown to form under certain conditions. Unlike the case of narrow superconductors, PSLs in wide films form on the background of a dynamic mixed state, when voltage already exists in the film due to the vortex flow. In view of this, a question arose as to the motion of vortices in the nonequilibrium region. We established[1] that the appearance of PSLs alters the dynamics of the vortices. This manifested itself clearly in the I–V curves in the region where a linear segment with a voltage-independent excess supercurrent appears. The differential resistance $R_d \sim R_N$ on this segment persisted as the magnetic field increased ($R_N$ is the resistance of the film in the normal state). What precisely happens to the moving vortices when PSL appear, however, remained an open question. How the vortex dynamics in the nonequilibrium region are affected by the normal component of the transport current $I_{tr}$ when vortices and PSL coexist, became a question of fundamental importance.[1,3] This is because vortex motion has hitherto been considered to be the result of the action, on the vortices, of Lorentz forces from the superconducting component $I_S$ of the current. At the same time, $I_S$ was assumed to be equal to the total current $I_{tr}$ flowing in the film (see, e.g., Ref. 4). However, in the vicinity of a PSL the current $I_{tr}$ is separated into the normal component $I_N$ and the superconducting component $I_S$ so that $I_{tr} = I_N + I_S$. In this paper, we present data, which show that, in the vicinity of a PSL, vortices interact only with the superconducting component of the transport current, and this determines the specific nature of the vortex flow in nonequilibrium regions.

The influence of PSLs on vortex flow can be studied using superconducting contacts S arranged at a distance $L$ ($\xi(T) < L < l_E$) from each other ($l_E$ is the penetration depth of the electric field into the superconductor). In the dynamic mixed state, until PSL appear, any pair of such contacts measures the voltage $V_V \sim d\chi/dt$ due to a change in the phase of the order parameter $\chi$ as vortices move between the contacts. The voltage jump $V_E$, which arises on the background of $V_V$ at the characteristic current $I^+$ and which is associated with the formation of a PSL, is measured only by that pair of S-contacts between which a PSL is formed. The voltage $V_E$ is zero across the potential contacts situated on one side of the PSL core.[1,5] Therefore, in a magnetic field these contacts will measure only the voltage caused by the vortex motion, and hence make it possible to monitor the change in the vortex flow pattern during the formation of PSL.

The procedures used to prepare the specimens and measure the *I–V* curves were similar to those described in Ref. 1. We studied tin films of thickness $d \sim 400$ Å and width $w \approx 50$ µm. We cut several potential leads ~5 µm apart each other in a ~30 µm-long film segment (the total length between the current leads was ~200 µm). For each segment of length $L \approx 5$ µm, we recorded the *I-V* curves in different magnetic fields and at several temperatures near $T_c$. The *I–V* curves of three segments are shown in Fig. 1. It is seen that the voltage jump and the linear part of $V(I)$ corresponding to the appearance of a PSL (Ref. 1), are observed in a zero magnetic field for $I > I^+$ only across segments 1 and 3, while $V = 0$ across segment 2 situated between them, in the same current range. The picture changes in a magnetic field. Until a PSL is formed at $I < I^+$, a dynamic mixed state is observed in all segments and $V = V_V \sim H v_L \sim d\chi/dt$ (Refs. 6,7) ($v_L$ is the vortex velocity). At currents above the characteristic current $I^+$, when PSLs form on segments 1 and 3, the voltage stops to change with the current on the segment 2 ($V = V_V =$ const). Similar *I–V* curves were obtained for several temperatures. The nonzero, current-independent voltage $V_V \sim v_L$ on segment 2 between two PSLs indicates that vortices move at a constant velocity here. Given the separation of the transport current into normal and superconducting components, $I_N$ and $I_S$, the constant vortex velocity after the formation of PSLs can be attributed to the fact that the vortices are solely acted upon by the current $I_S$. The later stays constant between two PSLs separated by a distance $\leq l_E$ with increasing the transport current (by analogy with phase-slip centers[2]). The interaction only with the superconducting component of the current is confirmed by our investigations of the vortex motion near an S–N boundary, where both $I_N$ and $I_S$ undergo spatial variation over a length $\sim l_E$ (Ref. 7, pp. 212-219).

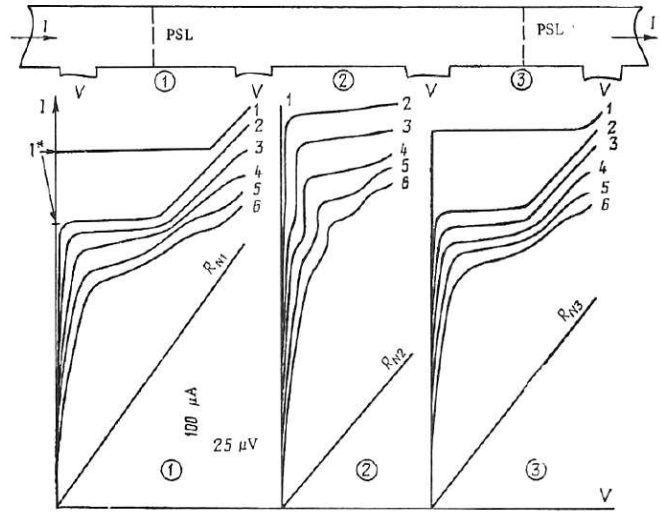

FIG. 1. *I-V* curves of three segments of a Sn film for $T/T_c = 0.974$ in different magnetic fields $H$(Oe): 1) 0, 2) 1.4, 3) 1.9, 4) 2.4, 5) 2.9, 6) 3.3. Top: schematic view of the specimen, the dashed line indicates the site of formation of a PSL; the length of each segment $L$ is 4.5-5.5 μm.

The *S–N* boundary was created between two segments of a tin film with different $T_c$ ($T_{c1} < T_{c2}$). The existence of the boundary was checked by the temperature dependence of the resistance, $R(T) \sim (1 - T/T_c)^{-1/4}$, in the range $T_{c1} \ldots T_{c2}$. The critical temperature of one part of the film was lowered, either by the evaporation of the normal metal onto half of the film or by variation of the film thickness ($d_2 < d_1$). In the last case, the *S–N* boundary could be obtained both in the range $T_{c1} \ldots T_{c2}$ and for $T < T_{c1}$, $T_{c2}$ by using a parallel magnetic field to destroy the superconductivity of the thicker part of the film. The film-under-study was divided into several segments near the *S–N* boundary by potential contacts (see inset in Fig. 2). The *I–V* curves of three segments in the S-region for different magnetic fields and the currents range $I_{tr} < I^+$ are given in Fig. 2.

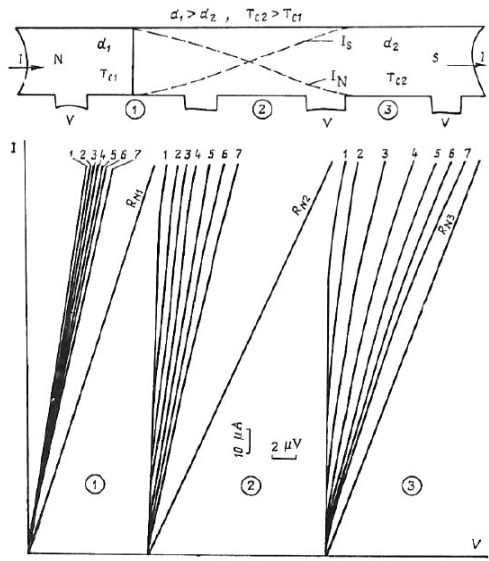

FIG. 2. *I-V* curves in magnetic fields, obtained on different segments of the film near the *S–N* boundary at $T_{c2} > T > T_{c1}$ in magnetic fields $H$(Oe): 1) 0, 2) 0.2, 3) 0.4, 4) 0.6, 5) 0.8, 6) 1.0, 7) 1.2.
Top: schematic view of the specimen; S and N are the superconducting and normal parts, respectively; film thickness $d_1 > d_2$; $T_{c2} > T_{c1}$; the dashed line represents the spatial dependence of $I_N$ and $I_S$ near the *S–N* boundary.

It is seen that for the same fields $H$ and currents $I$, the greater the distance between the segment (on which the *I–V* curve was taken) and the boundary, the higher the voltages $V \sim V_L$. This difference vanished when both parts of the film were superconducting. Moreover, the *I–V* curves of the individual segments are identical at distances greater than $l_E$ from the *S–N* boundary. Similar results were obtained when either of the methods mentioned above were used to form the *S–N* boundary.

Both measurements at currents below the characteristic current $I^+$ of a PSL formation and the increase in $V$ with rising magnetic field at the film parts far from the S–N boundary witness for the uniform vortex flow. This let us believe that the I-V characteristics of the film parts near the boundary (see Fig. 2) are determined by moving vortices. A voltage rise with increasing the distance from the S–N boundary at constant magnetic field and transport current is observed in the nonequilibrium region where the normal current component decays over the length $\sim l_E$ while the superconducting one increases up to the value $I_S = I_{tr}$. This makes it possible to explain the increase in $V$ and, therefore, in $v_L$ at areas more distant from the boundary by the action of the superconducting current component.

Thus, it is established that the dynamic mixed state persists in wide films after emergence of PSLs and thus nonequilibrium regions in their vicinity. However, since the transport current in a nonequilibrium region is divided in normal and superconducting components, the vortices interact only with the superconducting component of the current.[*]

Prior to the present experiments, it was not obvious that vortices move at constant velocity when a film is stratified in PSLs that corresponds to the region of the linear segment with $R_d \approx R_N$ in the I–V characteristic. The above data explicitly confirms this fact. Moreover, the dependence[1] of the excess current $I_0$, obtained by extrapolation of the linear segment in the I–V curve to the current axis, on the out-of-plane (perpendicular) magnetic field, $I_0 \sim (1-h)^2$ (here $h = H/H_{c2}$), now becomes clear.

In long films ($L \gg l_E$), where the number of PSLs is large, the I–V characteristic for $I_{tr} \gg I_c$ would have the form $V_E = (I_{tr} - I_{exc}) \cdot R_N$, by analogy with narrow superconducting channels[2], where the excess current $I_{exc} = A \cdot I_c$. Apparently, the dependence of $I_{exc}$ on the magnetic field $H$ is determined by the decoupling factor $\alpha \sim H$ (Ref. 8, pp.264-268), i.e., $I_{exc}(H) = I_{exc}(0)(1-h)$. As shown above, however, in a perpendicular field the voltage $V_E$ is augmented by a voltage $V_V$, of constant magnitude, caused by the vortex motion under the action of $I_{exc}$; $V_V = R_f / I_{exc}$, where $R_f$ is the flux-flow resistance. In this case, the experimental I–V characteristic in the region of the linear $V(I)$ dependence with a constant differential resistance $R_d \approx R_N$ can be written as

$$V = V_E + V_V = [I_{tr} - I_0(H)] R_N.$$

Using the expressions given above for $V_E$ and $V_V$, for the measured current $I_0(H)$ we get the experimentally observed dependence[1]

$$I_0(H) = I_{exc}(H)\ (1 - R_f/R_N) \sim I_{exc}(0)\ (1-h)^2.$$

It is assumed that $R_f \sim h$. In parallel fields, when vortices do not penetrate into the film $[d \ll \xi(T)]$, the decoupling factor $\alpha \sim H^2$, and $I_0(H) = I_{exc}(H) \sim (1-h^2)$ (Ref. 1).

The authors thank V. V. Shmidt for useful discussion of the results of the work.

---

[*] Independently of us, V.V. Schmidt and his colleagues came to the same conclusion. We take this opportunity to express our gratitude to V.V. Schmidt for familiarizing us with the results of the work before its publication.

[1] V. G. Volotskaya, I. M. Dmitrenko, and A. G. Sivakov, Fiz. Nizk. Temp. 10, 347 (1984) [Sov. J. Low Temp. Phys. 10, 179 (1984)].

[2] B. I. Ivlev and N. B. Kopnin, Usp. Fiz. Nauk 142, 435 (1984) [Sov. Phys. Usp. 27, 206 (1984)].

[3] V. Yu. Taranenko and A. I. D'yachenko, in: Abstracts of Twenty-Third All-Union Conference on Low-Temperature Physics (Tallin, October 23-25, 1984] [in Russian], Tallin (1984), Part 1, C-47, pp. 176-177.

[4] L. P. Gor'kov and N. B. Kopnin, Usp. Fiz. Nauk 116, 413 (1975) [Sov. Phys.Usp. 18, 496 (1975)].

[5] G. J. Dolan and L. D. Jackel, Phys. Rev. Lett. 39, 1628 (1977).

[6] I. O. Kulik, Preprint No. 99- 66 Dep. Physicotechnical Institute of Low Temperatures, Academy of Sciences of the Ukrainian SSR, Kharkov (1966).

[7] V. V. Shmidt, Introduction to the Physics of Superconductors [in Russian]. Nauka, Moscow (1982).

[8] M. Tinkham, Introduction to Superconductivity, Krieger, Malabar, Florida (1980) [reprint of McGraw-Hill, N.-Y. (1975)], Ch. 8.

Translated by Eugene Lepa (1985), revised by Oleg Turutanov (2025).